\documentclass[a4paper,10pt,twoside]{cpc-hepnp}

\usepackage{multicol}
\usepackage{graphicx}
\usepackage{booktabs}
\usepackage{amssymb,bm,mathrsfs,bbm,amscd}
\usepackage[tbtags]{amsmath}
\usepackage{lastpage}

\begin{document}



\title{Chiral symmetry restoration in excited hadrons and dense matter }

\author{%
L. Ya. Glozman$^{1)}$\email{leonid.glozman@uni-graz.at}%
}
\maketitle

\address{  Institute for Physics, Theoretical Physics Branch, 
University of Graz,Universit\"atsplatz 5, A-8010, Graz, Austria\\}

\begin{abstract}
We overview two interconnected topics: possible effective restoration of chiral
symmetry in highly excited hadrons and possible existence of confined but chirally symmetric
matter at low temperatures and high densities.
\end{abstract}



\begin{multicols}{2}

\section{Introduction}

The question of mass generation and the related question of interconnections
of confinement and chiral symmetry breaking are central for QCD. In order to
answer these questions we have to understand the gross structure of the hadron
spectrum in the light quark sector and correlate it with  interactions of hadrons
with the Nambu-Goldstone bosons of broken chiral symmetry. Given symmetry patterns
observed in the hadron spectra \cite{G1,C1}, their correlations with the couplings to 
pions \cite{G2}, or,
more generally, with their axial properties, one can obtain the insight into the principal
mechanism responsible for the mass generation, whether or not the mass of hadrons is directly
related to the quark condensate of the vacuum. There are strong hints that in the highly
excited hadrons the physics of mass generation is mostly unrelated to spontaneous breaking
of chiral symmetry in the vacuum, i.e.  most part of the hadron mass is not due to
the quark condensate of the vacuum. This phenomenon, if correct, is referred to as
effective restoration of chiral symmetry.  This is just in contrast to physics of the lowest
lying hadrons like nucleon, where   mass is mostly induced by the quark condensate.

The issues of mass generation and interconnections of confinement and chiral symmetry are
critically important for our view of the QCD phase diagram. For many years it was believed
that in the confining mode  chiral symmetry should be strongly broken. This is certainly
true in the vacuum, as it follows from the model-independent 't Hooft anomaly matching
conditions \cite{amc}. Extrapolating these constraints to the nonzero temperature
and density regions, one naively concludes that there cannot be a phase in the QCD phase diagram
that is confining but with vanishing quark condensate. This picture was supported by 
simple models of confinement and chiral symmetry breaking. It is this argument which was a basis
for a belief that the deconfinement and chiral restoration phase transitions coincide 
in the temperature - chemical potential plain. Then, given this apriori belief, the QCD phase
diagram was modeled after the Nambu and Jona-Lasinio model phase diagram (or variants of thereof). This
model is nonconfining.

In the large
$N_c$ limit QCD is confining at low temperatures up to arbitrary large density and such
a matter was called quarkyonic \cite{MP}. Then, at some reasonably large density
one can obtain a confining but chirally symmetric phase at low temperatures \cite{GW1,GW2,GW3}.
In such a phase the standard quark-antiquark condensate of the QCD vacuum vanishes
and chiral symmetry is restored (or it can be  slightly broken via the chiral
breaking phenomena near the Fermi surface  - the chiral density waves \cite{cdw}).
The important point is that the standard
quark condensate of the vacuum vanishes at high density and the bulk mass 
in the confining mode has mostly the
chirally symmetric origin.

\section{Chiral symmetry breaking and its implications }  

 The $SU(2)_L \times SU(2)_R$ axial symmetry of the QCD Lagrangian in the chiral limit
is dynamically broken in the vacuum. Then there appear massless Goldstone bosons
associated with broken axial symmetry. 

Another direct evidence of dynamical chiral symmetry breaking in the vacuum is
absence of the chiral parity doublets in the observed low-lying spectrum. If there is not
a chiral partner to the nucleon, then its mass is generated  through the quark condensate
of the vacuum. Such a behavior can be modeled within the QCD sum rule approach,
within the linear sigma-model,  within the NJL model, variants of the
bag model, constituent quark model with very massive constituent quarks, or within the
Skyrme model.
One cannot exclude, however, that some small part of nucleon mass is
not related to the quark condensate. The (partial) axial vector current conservation
translates this mass, via the Goldberger-Treiman relation, to the pion-nucleon coupling
constant. Hence, the large pion-nucleon coupling constant encodes the physical origin
of the nucleon mass as due to chiral symmetry breaking in the vacuum. It can be used
as a natural measure for chiral symmetry breaking effect in a hadron.

\section{Effective chiral restoration in baryon spectra} 

The nucleon excitation spectrum exhibits obvious patterns of parity
doublets, see Fig. 1. Similar patterns are seen in the Delta spectrum. 
The linear axial transformation in the isospin space (i.e. the chiral
$SU(2)_L \times SU(2)_R$ transformation) mixes the nucleon (or delta) states
of a given parity with the nucleon or delta states of opposite parity.
Then unbroken chiral symmetry requires existence of parity doublets in
the nucleon and delta spectra that are not connected to each other,
or quartets, i.e. parity doublets in the nucleon and delta spectra that
are members of same higher representation. The absence of parity
doublets in the low-lying spectrum is an evidence of chiral symmetry breaking
in the vacuum. Appearance of parity doublets in highly excited nucleons and deltas
was taken as evidence of effective chiral symmetry
restoration \cite{G3,CG1}. Of course, to claim a general pattern
one needs a discovery of the still missing partners to the well
established states with $7/2^-$ and   $11/2^-$.

While these parity  doublets are impressive, by themselves they
are only suggestive, because they could be  accidental. Given a statistical 
analysis of ref. \cite{JPS1} the latter is unlikely.
However, there could other reason, not related to chiral symmetry,
responsible for the doubling. 
Then we need other  observables that are sensitive specifically to
chiral symmetry and that would correlate
with the observed  degeneracies. Such  observables are axial properties
of states: their axial charges and couplings to pions. 

\begin{center}
\includegraphics[width=5cm,angle=-90,clip=]{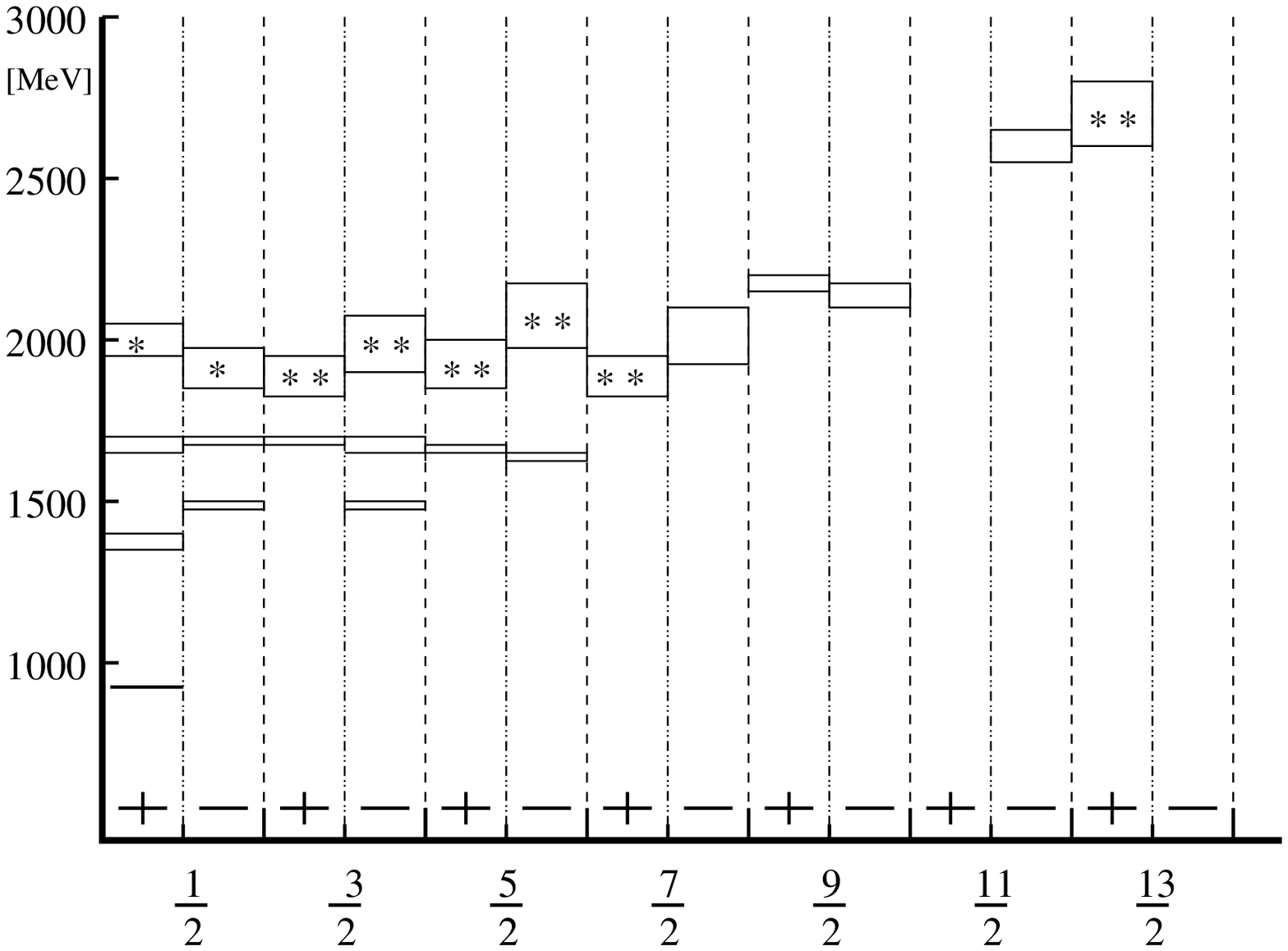}
\figcaption{\label{fig1}    Low- and high-lying nucleons. Those states which are
not yet established are marked by ** or * according to the PDG classification.}
\end{center}

Assume that the parity doublets are accidental and they are degenerate 
due to some other
reason, not related to chiral symmetry.  This means that
these states are not chiral partners. Then their mass  is
 induced by the quark condensate of the vacuum, like
 nucleon mass. The axial charges of these states should be
expected to be of order 1. The Goldberger-Treiman relation 
then tells  that these states must be very strongly coupled to pions
and that the coupling constant to  pion should be comparable with the
pion-nucleon coupling constant. Such states
should have a large decay coupling to the $\pi N$ channel. In contrast,
the effective chiral restoration requires that the axial charges of
these states should be small (as compared to the nucleon axial charge)
and they must have small decay coupling constants into the $\pi N$ channel
\cite{G2,JPS2}. The diagonal axial charges of excited states as well as their
diagonal couplings to pions is difficult
to extract from  experiment. The decay coupling constants can be obtained
from the known decay widths, however.  These decay coupling constants 
in units of the well-known pion-nucleon coupling constant are
shown in Table 1. One clearly observes that all those excited nucleons that
are assumed from the spectroscopic patterns to be in approximate chiral
multiplets have a very small decay coupling constant into the $\pi N$ channel.
In contrast, the only well established excited nucleon, $ N_{{3/2}^-}(1520)$, in
which case a chiral partner cannot be identified from the spectrum, has
a very large $\pi N$ decay coupling constant. Consequently its mass origin,
 should be the quark condensate of the vacuum, similar to nucleon.
It is a very interesting question what dynamics makes this state so peculiar.
One observes a 100\% correlation of the spectroscopic patterns with the
 $\pi N$ decays as predicted by effective chiral restoration. It is noteworthy
 that a specific small value of the decay coupling constant cannot be predicted
 by chiral symmetry and depends on the  microscopic structure of the state. The 
 approximate chiral symmetry only requires  
 that these decay constants must be small.

\end{multicols}
\begin{center}
\tabcaption{ \label{tab1} Chiral multiplets of excited nucleons.
Comment:  There
are two possibilities to assign the chiral representation:
$(1/2,0) \oplus (0,1/2)$ or $(1/2,1) \oplus (1,1/2)$ because
there is a possible chiral pair in the $\Delta$ spectrum
with the same spin with similar mass. }
\vspace{-3mm}
\footnotesize
\begin{tabular*}{170mm}{@{\extracolsep{\fill}}cccc} 

\toprule Spin & Chiral multiplet &  Representation  & 
$(f_{B_+N\pi}/f_{NN\pi})^2 -  (f_{B_-N\pi}/f_{NN\pi})^2$ \\ \hline
1/2& $N_+(1440 ) - N_-(1535)$ & $(1/2,0) \oplus (0,1/2)$ &
   0.15 - 0.026    \\
\hline

1/2& $N_+(1710) - N_-(1650)$ & $(1/2,0) \oplus (0,1/2)$ &
 0.0030 - 0.026   \\

3/2& $N_+(1720) - N_-(1700)$ & $(1/2,0) \oplus (0,1/2)$ &
 0.023 - 0.13     \\

5/2&$N_+(1680) - N_-(1675)$ & $(1/2,0) \oplus (0,1/2)$ &
 0.18 - 0.012   \\

7/2&$N_+(?) - N_-(2190)$ &  see comment   &
  ? - 0.00053   \\

9/2&$N_+(2220) - N_-(2250)$ &
 see comment  &
 0.000022 - 0.0000020  \\

11/2&$N_+(?) - N_-(2600)$ &   see comment  &
 ? - 0.000000064   \\

\hline
\hline
3/2& $ N_-(1520)$ & no chiral partner &
   2.5     \\

\bottomrule
\end{tabular*}%
\end{center}

\begin{multicols}{2}
 
  The diagonal axial charges of excited states cannot be measured experimentally,
 but in principle can be obtained in lattice simulations. On the lattice it is an
 intrinsically difficult problem to extract the highly excited states. Nevertheless
 a progress has been achieved in identification of the lowest negative parity
 excitations \cite{BURCH}. Recently  first lattice results
 for the diagonal axial charges of the states  $ N_{{1/2}^-}(1535)$ and  
 $ N_{{1/2}^-}(1650)$ have appeared \cite{TK}. These first  results are limited to
 rather large quark masses and also require confirmation by other groups. 
 Assuming a naive extrapolation of the results to the physical point one
 concludes that the axial charge of the lowest negative parity excitation,
$ N_{{1/2}^-}(1535)$, is very small. This is  consistent with a possible
identification of the $N_{{1/2}^+}(1440) -  N_{{1/2}^-}(1535)$ pair as the
lowest chiral pair. The chiral symmetry breaking effects are
still large in this case (because of rather large splitting of the states).
 May be this also explains a long-standing puzzle
why the Roper state has so small mass. Given  large error bars for
the next excited state,  $ N_{{1/2}^-}(1650)$, it is difficult to extrapolate
its axial charge to the physical point. One should mention that the obtained values of
the axial charges are also consistent with the $SU(6)_{FS} \times O(3)$ quark
model prediction assuming that there is not mixing of the $S=1/2$ state
$ N_{{1/2}^-}(1535)$ with the  $S=3/2$ state
$ N_{{1/2}^-}(1650)$ via the tensor quark-quark force \cite{N,GN1,P}. Within the
Isgur-Karl type quark models such a mixing is very strong; it makes the
axial charge of $ N_{{1/2}^-}(1535)$ to be of the order 1. Such a strong mixing
is also very important to obtain a reasonable   fit of strong baryon decays
within the constituent quark model.

Here one more comment on 
the quark (or large $N_c$)
description of baryon decays is relevant. These models operate with the
nonrelativistic decay amplitudes with improper nonrelativistic phase space factors. 
They try to fit, with some free parameters, decay widths, rather than the coupling 
constants. Physics is contained in the coupling constants, however. Then a proper procedure
would be to compare the quark model  predictions with the coupling
constants.   

\section{Highly excited mesons}

Fig. 2 shows the spectra of the well established mesons from the PDG and new,
not yet confirmed $\bar n n$ states from the partial wave analysis \cite{AN,B}
of $\bar p p$ annihilation at LEAR (CERN). Obvious large degeneracy of the
high-lying mesons is seen. How could we understand such a degeneracy? 
 This data have been analyzed in
refs. \cite{G4,G5} and it was shown that the degeneracies of the high-lying
states with $J=0-3$ are consistent with the conjecture of effective 
$SU(2)_L \times SU(2)_R$ and $U(1)_A$ restorations. A prediction was
made that the pattern for the high-lying $J=4$ states should be similar
to the pattern of $J=2$ states and the pattern for the $J=5$ mesons should
be the same as the pattern for the $J=3$ mesons. 
There are $a_4$ and $f_4$ positive parity states in the band around 2 GeV
 and their possible
partners of opposite parity are missing. Similar happens with the $J=5$
states in the 2.3 GeV band. The absence of the chiral partners for these
highest spin mesons would be a difficulty for the chiral restoration scenario. 
Consequently
a key question is whether these states do not exist or they cannot be
seen in the $\bar p p$ annihilation, even if they exist. 
It turns out that the latter is correct \cite{GS}.

\end{multicols}
\begin{center}
\includegraphics[width=12cm]{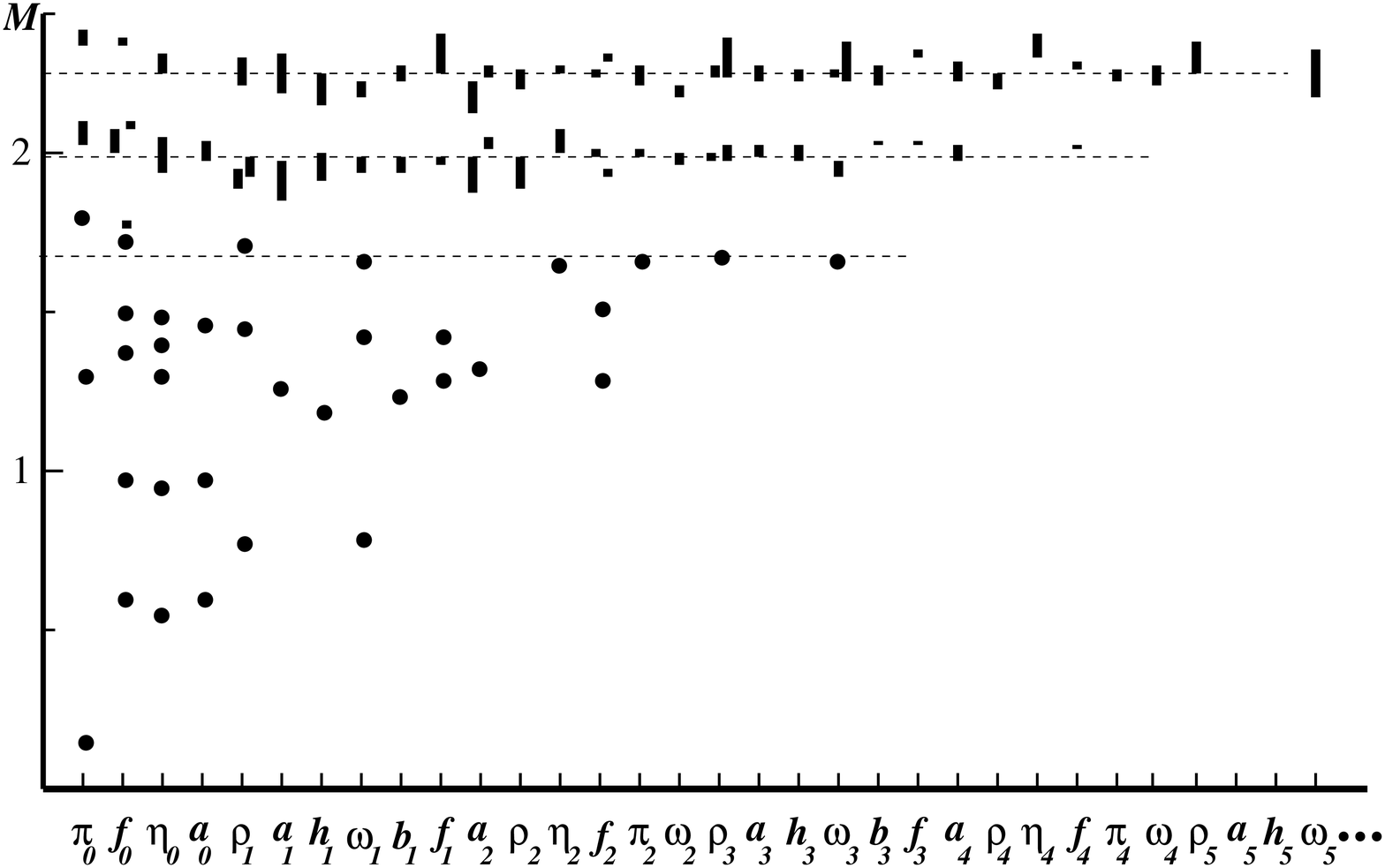}
\figcaption{\label{fig2}   Masses (in GeV) of the well established  states from PDG 
(circles) and 
new $\bar n n$
states   from the proton-antiproton annihilation (strips). Note
that the well-established states include $f_0(1500), f_0(1710)$, which
are the glueball and $\bar s s$ states with some mixing and hence are
irrelevant from the chiral symmetry point of view. Similar, the
 $f_0(980), a_0(980)$ mesons most probably are not $\bar n n$ states and
 also should be excluded from the consideration. The same is true for
 $\eta(1475)$, which is the $\bar s s$ state and  $\eta(1405)$ with
 the unknown nature. }
\end{center}
\begin{multicols}{2}

Consider, for example, the missing $J=4$ states of negative parity in the 2 GeV band. They 
{\it all} require the $L=4$ partial wave in the $\bar p p$ system. In contrast,
the observed $a_4$ and $f_4$  mesons are produced in the $L=3$ partial wave.
From Fig. 2 it follows that the missing chiral partners should be expected
with mass $ 2000 \pm 50$ MeV. At such energy the $L=4$ partial wave in the 
$\bar p p$ system is suppressed as compared to the $L=3$ partial wave
 by the factor $10^1 - 10^3$
by the centrifugal repulsion in the  $\bar p p$ system.
A signal from this  missing states is very weak as compared to the
observed $a_4$ and $f_4$ states. The same suppression factor is valid with
respect to the negative parity states {\it seen} in the 2.3 GeV band. With such a weak
signal the $\chi^2$ fit cannot reveal these missing states, even if they exist.
Similar analysis can be done for the $J=5$ states in the 2.3 GeV band. One then
concludes that the existing experimental data \cite{AN,B} on highly excited mesons
cannot answer a question about existence or non-existence of these missing states
and new types
of experiments with polarization are required to answer this conceptually important
question.

The chiral and $U(1)_A$ symmetries cannot explain degeneracies
of the states with different spins. Such a degeneracy can be obtained, however, if one
assumes a principal quantum number $n + J$ on top of chiral restoration \cite{GN2}. 

There
exists an alternative conjecture about  nature of the large degeneracy. If
 these high-lying states behaved nonrelativistically (i.e., the
valence quarks were nonrelativistic)  the degeneracy could be obtained assuming 
the standard nonrelativistic
$LS$ coupling scheme and  a principal quantum number $n + L$,
where $L$ is the orbital angular momentum of quarks \cite{A,K,SV}. In the nonrelativistic quantum
mechanics $L$ can indeed be a good quantum number in absence of the
spin-orbit force (compare this, e.g., with the nonrelativistic Hydrogen atom).

Such a scenario is inconsistent with two basic facts. In QCD, which is a highly
relativistic quantum field theory, there is only one conserved
angular momentum, $J$. The principal quantum number $n + L$
would imply that there are three independent conserved angular momenta, $L,S,J$!
Such an assumption is also inconsistent with the stringy picture, on which it is
based. The ends of the rotating string move at the speed of light. Then
the quarks at the ends of the string are ultrarelativistic and must have a definite chirality because
only chiral quarks can move at the speed of light. The stringy
picture does imply the {\it unbroken} chiral symmetry and would in fact predict the
missing states in the 2 and 2.3 GeV bands. A consistent relativistic string model
with quarks at the ends of the string is an open issue.

Another interesting question is a linearity of the Regge trajectories.
The
leading nucleon angular Regge trajectory is approximately linear, a well known
fact. The daughter trajectory is highly non-linear, however. While the
high-spin states $J=5/2$ and $J=9/2$ of both positive and negative
parity are approximately degenerate, there is not a degenerate state
of the opposite parity for the nucleon. In the meson spectrum the leading
angular Regge trajectory is also approximately linear. Should one expect a 
linearity of all daughter Regge trajectories in the meson spectrum? 

\section{Models}

One cannot solve QCD, even in the large $N_c$ limit. Hence at the
moment the only useful tool to address the problem of highly excited
hadrons is modeling.
The model must contain all principal elements of QCD that are relevant to
the present problem. It must be (i) relativistic and field-theoretical in nature,
(ii) chirally symmetric, (iii) confining, (iv) it should provide dynamical
breaking of chiral symmetry. It is highly nontrivial to meet all these
requirements within one and the same model. For example the NJL model (or
variants of thereof) is chirally symmetric and guarantees spontaneous breaking
of chiral symmetry. It is not confining, however. In contrast, the potential
constituent quark models do not respect points (i), (ii) and (iv).

There exists such a model, however \cite{Ya,AD}. The model is a generalization of
the 't Hooft model \cite{h}. The 't Hooft model is QCD in large $N_c$ limit
in 1+1 dimensions. Due to its low-dimensional nature it is an {\it exactly}
solvable field theory. One can analytically calculate all required quantities:
the quark condensate, the meson spectrum, etc. It is useful to understand
{\it how} this field theory is solved. In 1+1 dimensions the highly nonlinear
gluodynamics is exactly reduced to the Coulomb interaction alone, which is 
an instantaneous linear potential of the Lorentz-vector type. To address
the problem of dynamical chiral symmetry breaking one has to solve the gap 
(Schwinger-Dyson) equation. Given the quark Green function obtained from the
gap equation, it is possible to solve the Bethe-Salpeter equation for mesons,
etc. However, in 1+1 dimensions a rotational motion and angular momenta
are absent, that are crucial for effective chiral restoration. The model  
\cite{Ya,AD} is a straightforward generalization of the 't Hooft model to
3+1 dimensions. It is postulated within the model that there exists  a
linear instantaneous Coulomb-type potential in 3+1 dimensions. All other possible
gluonic interactions are neglected. Given its "simplicity", the model can be solved
numerically.
 The problem of chiral restoration in excited mesons
has been addressed in ref. \cite{WG}. A complete spectrum of mesons has been
calculated and a fast effective chiral restoration with increasing  spin $J$ has
been demonstrated. It is instructive to outline a physical mechanism responsible
for the phenomenon. When one increases the spin of a hadron $J$, one also increases 
a typical momentum of valence quarks.
The chiral symmetry breaking dynamical mass of quarks is important only at
low momenta. At large $J$ all low momenta components are suppressed 
in the meson wave function by the centrifugal repulsion and consequently the chiral 
symmetry breaking dynamical mass of quarks
gets irrelevant. Consequently one observes the effective chiral restoration.  Due to its simplicity, 
the model 
cannot reproduce a degeneracy of chiral multiplets with different spins \cite{B1}.
It is much more difficult to solve the model  for baryons. Nevertheless some steps
have been done \cite{NSR,BL} and similar effective chiral restoration with increasing
$J$ has been observed. 

The problem of highly excited hadrons has been addressed for the last years
within many different holographic models. All existing holographic models
of hadrons suffer an essential disease, however. The holographic models are assumed to satisfy the
AdS/CFT matching conditions at the ultraviolet border, where chiral symmetry
is not broken. It has been recently proven \cite{GN3} that in reality they do {\it not}
satisfy these matching conditions. As such, they are not suitable to address
the problems related to interconnections of confinement and chiral symmetry
in hadrons.

\section{Chiral restoration in the quarkyonic matter}

Quite recently McLerran and Pisarski  suggested  a
new state of the matter - the quarkyonic matter \cite{MP}. Their crucial
observation was that in the large $N_c$ limit at low and moderate temperatures,
confinement persists up to arbitrary high densities. There are no dynamical quark loops and
hence nothing screens the confining gluon propagator. 

At some critical density one should expect a chiral phase transition, namely
one expects that the standard quark-antiquark condensate of the QCD vacuum should
vanish. Then one arrives at a subphase within the quarkyonic matter
that is manifestly confining and at the same time the quarks condensate vanishes.
How could it be ?!  This question was addressed 
in refs. \cite{GW1,GW2,GW3}. The same confining and chirally symmetric model was
 chosen that had been used for study of effective chiral restoration in excited
hadrons. The following mechanism for confining matter with vanishing quark-antiquark
condensate was observed. The quark Green function, that is a solution of
the gap equation, acquires not only the chiral symmetry breaking Lorentz-scalar
part but also the chirally symmetric Lorentz-vector part. Both these parts are
infrared-divergent, which guarantees that there are no single quarks in the spectrum.
The infrared divergence cancels exactly in all color-singlet quantities, such us
the quark condensate or hadronic excitations. At some critical density 
the Lorentz-scalar part of the quark self-energy as well as the quark
condensate of the vacuum vanishes and one observes a chiral phase transition. This
happens exclusively due to the Pauli blocking of the quark levels that are required for existence
of the quark condensate. The Lorentz-vector part of the quark self-energy does not vanish, 
however, and is still infrared-divergent. This guarantees that
a single quark does not exist even above the phase transition.
This is just in contrast to the nonconfining NJL model (or variants of thereof) where the Lorentz-vector
part of the quark self-energy is absent and one inevitably has a massless free quark
in the chirally symmetric regime. The infrared singularity cancels exactly in any
possible color-singlet excitation of the matter and such an excitation has a finite
well-defined energy. Consequently even in the absence of the chiral symmetry breaking
via the quark-antiquark condensate one obtains a confining matter with color-singlet
excitations only. The mass of a dense confining
matter is not related to the chiral symmetry breaking via the quark condensate.

It does not mean, however, that such a matter with vanishing quark condensate will be
exactly chirally symmetric. It  can have a small amount of chiral symmetry breaking
via the chiral density waves near the Fermi surface \cite{cdw}. But in any case
the bulk of the mass of this confining matter is not related to chiral symmetry breaking.

\acknowledgments{The author acknowledges support of the Austrian Science Fund
through grants P19168-N16; P21970-N16}

\end{multicols}

\vspace{-2mm}
\centerline{\rule{80mm}{0.1pt}}
\vspace{2mm}

\begin{multicols}{2}

\end{multicols}

\clearpage

\end{document}